\let\includefigures=\iffalse
%
\let\useblackboard=\iftrue
%
%
\newfam\black
\input harvmac
\includefigures
\message{If you do not have epsf.tex (to include figures),}
\message{change the option at the top of the tex file.}
\input epsf
\def\figin{\epsfcheck\figin}\def\figins{\epsfcheck\figins}
\def\epsfcheck{\ifx\epsfbox\UnDeFiNeD
\message{(NO epsf.tex, FIGURES WILL BE IGNORED)}
\gdef\figin##1{\vskip2in}\gdef\figins##1{\hskip.5in}
\else\message{(FIGURES WILL BE INCLUDED)}%
\gdef\figin##1{##1}\gdef\figins##1{##1}\fi}
\def\DefWarn#1{}
\def\figinsert{\goodbreak\midinsert}
\def\ifig#1#2#3{\DefWarn#1\xdef#1{fig.~\the\figno}
\writedef{#1\leftbracket fig.\noexpand~\the\figno}%
\figinsert\figin{\centerline{#3}}\medskip\centerline{\vbox{\baselineskip12pt
\advance\hsize by -1truein\noindent\footnotefont{\bf Fig.~\the\figno:} #2}}
\bigskip\endinsert\global\advance\figno by1}
\else
\def\ifig#1#2#3{\xdef#1{fig.~\the\figno}
\writedef{#1\leftbracket fig.\noexpand~\the\figno}%
\global\advance\figno by1}
\fi
\useblackboard
\message{If you do not have msbm (blackboard bold) fonts,}
\message{change the option at the top of the tex file.}
\font\blackboard=msbm10 scaled \magstep1
\font\blackboards=msbm7
\font\blackboardss=msbm5
\textfont\black=\blackboard
\scriptfont\black=\blackboards
\scriptscriptfont\black=\blackboardss

\else

\fi
%
\def\yboxit#1#2{\vbox{\hrule height #1 \hbox{\vrule width #1
\vbox{#2}\vrule width #1 }\hrule height #1 }}
\def\fillbox#1{\hbox to #1{\vbox to #1{\vfil}\hfil}}
\def\ybox{{\lower 1.3pt \yboxit{0.4pt}{\fillbox{8pt}}\hskip-0.2pt}}

\def\comments#1{}

\def\tr{{\rm Tr\ }}

\def\CA{{\cal A}}

\def\CM{{\cal M}}

\def\CL{{\cal L}}

\def\nl{\hfill\break}

\def\II{\relax{I\kern-.10em I}}
\def\IIa{{\II}a}

\def\IZ{\relax\ifmmode\mathchoice
{\hbox{\cmss Z\kern-.4em Z}}{\hbox{\cmss Z\kern-.4em Z}}
{\lower.9pt\hbox{\cmsss Z\kern-.4em Z}}
{\lower1.2pt\hbox{\cmsss Z\kern-.4em Z}}\else{\cmss Z\kern-.4em
Z}\fi}
\def\IB{\relax{\rm I\kern-.18em B}}
\def\IC{{\relax\hbox{$\inbar\kern-.3em{\rm C}$}}}
\def\ID{\relax{\rm I\kern-.18em D}}
\def\IE{\relax{\rm I\kern-.18em E}}
\def\IF{\relax{\rm I\kern-.18em F}}
\def\IG{\relax\hbox{$\inbar\kern-.3em{\rm G}$}}
\def\IGa{\relax\hbox{${\rm I}\kern-.18em\Gamma$}}
\def\IH{\relax{\rm I\kern-.18em H}}
\def\II{\relax{\rm I\kern-.18em I}}
\def\IK{\relax{\rm I\kern-.18em K}}
\def\IP{\relax{\rm I\kern-.18em P}}

%

\def\inbar{\,\vrule height1.5ex width.4pt depth0pt}

\font\cmss=cmss10 \font\cmsss=cmss10 at 7pt
\def\IR{\relax{\rm I\kern-.18em R}}

\def\BR{\IR}

\def\BR{\IR}
\def\BC{\IC}

\def\lp10{l_P^{10}}
\def\lp11{l_P^{11}}
\def\R11{R_{11}}

\Title{\vbox{\baselineskip12pt\hbox{hep-th/9712230}
\hbox{RU-97-104}}}
{\vbox{
\centerline{Fractional Branes and Wrapped Branes} }}
\centerline{Duiliu-Emanuel Diaconescu$^1$, Michael R. Douglas$^{1,2}$
and Jaume Gomis$^1$}
\medskip
\centerline{$^1$Department of Physics and Astronomy}
\centerline{Rutgers University }
\centerline{Piscataway, NJ 08855--0849}
\medskip\centerline{$^2$ Institut des Hautes \'Etudes Scientifiques}
\centerline{Le Bois-Marie, Bures-sur-Yvette, 91440 France}
\centerline{\tt duiliu, mrd, jaume@physics.rutgers.edu}
\bigskip
\noindent
We discuss the ``fractional D-branes'' which arise
in orbifold resolution.  We argue that they arise as
subsectors of the Coulomb branch of the quiver gauge theory used
to describe both string theory D-brane and Matrix theory on an orbifold,
and thus must form part of the full physical Hilbert space.
We make further observations confirming their interpretation as
wrapped membranes.

\Date{December 1997}
%
\lref\dlp{J.~Dai, R.~G.~Leigh and J.~Polchinski,
``New connections between string theories,'' Mod. Phys. Lett. {\bf A4}
(1989) 2073;
J.~Polchinski, ``Dirichlet branes and Ramond-Ramond charges,''
Phys.~Rev.~Lett.~{\bf 75} (1995) 4724-4727;
hep-th/9510017.}
\lref\kron{P. Kronheimer, J. Diff. Geom. {\bf 28}1989)665.}
\lref\Ab{P.S. Aspinwall, TASI 96, hep-th/9611137.}
\lref\leigh{R. Leigh, ``Dirac-Born-Infeld action from Dirichlet sigma
model,'' Mod. Phys. Lett. {\bf A4} (1989) 2767.}
\lref\witten{E. Witten, Nucl. Phys. B443 (1995) 85; hep-th/9503124.}
\lref\dos{M. R. Douglas, H. Ooguri and S. H. Shenker,
Phys. Lett. {\bf B402} (1997) 36-42; hep-th/9702203.}
\lref\dstrings{M. R. Douglas, ``D-branes in Curved Space,''
lecture at Strings '97.}
\lref\dcurve{M. R. Douglas, ``D-branes in Curved Space,''
hep-th/9702048.}
\lref\dm{M. R. Douglas and G. Moore, hep-th/9603167.}
\lref\dgm{M. R. Douglas, B. Greene and D. R. Morrison,
Nucl. Phys. {\bf B506} (1997) 84, hep-th/9704151.}
\lref\polpro{J. Polchinski, Phys.Rev. {\bf D55} (1997) 6423, hep-th/9606165.}
\lref\egs{M. R. Douglas, hep-th/9612126.}
\lref\bfss{T. Banks, W. Fischler, S. H. Shenker and L. Susskind,
Phys.Rev. {\bf D55} (1997) 6382, hep-th/9610043.}
\lref\dkps{M. R. Douglas, D. Kabat, P. Pouliot and S. Shenker,
Nucl.Phys. {\bf B485} (1997) 85, hep-th/9608024.}
\lref\dg{D.-E. Diaconescu and J. Gomis, hep-th/9707019.}
\lref\senseiberg{A. Sen,
hep-th/9709220; \nl
N. Seiberg,  Phys.Lett. {\bf B402}
(1997) 36, hep-th/9710009.}
\lref\hard{M. R. Douglas and H. Ooguri, hep-th/9710178.}
\lref\kp{D. Kabat and P. Pouliot, Phys.Lett. {\bf B402} (1997) 36,
hep-th/9603127; \nl
U.~H.~Danielsson, G.~Ferretti, and B.~Sundborg, Phys.Lett. {\bf B402} (1997)
36, hep-th/9603081.}
\lref\johnmy{C. Johnson and R. Myers, Phys.Rev. {\bf D55} (1997) 6382,
hep-th/9610140.}
\lref\hm{J. Harvey and G. Moore, hep-th/9609017.}
\lref\wittena{E. Witten,
hep-th/9707093.}
\newsec{Introduction}

D-branes propagating on resolved orbifolds provide a simple and tractable
example of how geometry at short distances
arises as the low energy configuration space
of a world-volume gauge theory \refs{\dm,\dkps,\polpro,\johnmy,\dgm}.
They also provide a simple and explicit starting point for the definition
of Matrix theory \bfss\ on these spaces \refs{\egs,\dos,\dg,\hard}.

The construction (reviewed in section 2) starts with maximally supersymmetric
gauge theory, to which a projection acting simultaneously
on space-time and Chan-Paton indices is applied,
leading to a `quiver' gauge theory whose
Higgs branch is the orbifold.  In string theory, the twist fields which
parameterize the blow-up of the orbifold control Fayet-Iliopoulos
terms, which resolve the low energy configuration space into a smooth space.

In \polpro\ it was pointed out that these theories, with zero
Fayet-Iliopoulos terms, also have a Coulomb branch.
This was interpreted as describing {\it wrapped} D-branes --
if the original theory described D$p$-branes, the Coulomb branch
describes a collection of D$p+2$-branes wrapped about the various
two-cycles which have shrunk to zero volume in the orbifold limit.
The simplest reason to believe this is the following.
The (classical) Coulomb branch is parameterized by scalar
expectation values describing coordinates in the space transverse
to the orbifold, and it has larger unbroken gauge
symmetry: instead of $U(1)$ (for a single D-brane) it has unbroken
$U(1)^n$ (for the $A_{n-1}$ singularity with $n-1$ two-cycles).
These parameters and gauge symmetry are consistent with an
interpretation as $n$ objects each bound to the fixed point in the
orbifold, but free to move in the transverse dimensions.

In string theory, this interpretation can be confirmed by showing
that the various objects are sources of the appropriate twisted
sector RR fields, using the techniques of \dm.
We describe this computation in section 2.
{}From its form, it is clear that one can treat the objects independently
by keeping an appropriate subsector of the gauge theory (e.g. a
single $U(1)$), and that
each has mass $1/n$ of the original D-brane,
motivating the name `fractional branes.'
These masses agree with string
theory expectations, arising from non-zero
$\int B$ on the cycles.

{}From the space-time point of view, the wrapped membranes are the
gauge bosons of ADE enhanced gauge symmetry (for $\BC^2/\Gamma$)
given masses by gauge symmetry breaking.  In perturbative
string theory, the
gauge symmetry is always broken by $\int B$.
In terms of the dual heterotic string theory,
this can be thought of as a Wilson line $A_{11}$.
One can also modify the
gauge symmetry breaking by turning on additional Wilson lines;
these correspond to turning on the FI terms. This will be shown more
explicitly  in section 3.

Turning on FI terms will change the masses of the gauge bosons,
but they will still exist as BPS states in the theory.
On the other hand, from the D-brane gauge theory point of view,
the FI terms lift the Coulomb branch and remove these supersymmetric
vacua!
How can we reconcile these two statements?

The basic resolution is the observation that the Coulomb
branch describes a state containing several charged gauge bosons,
whose charges sum to zero.  Such a state generally will not be BPS
and thus the corresponding
gauge theory state will not be a supersymmetric vacuum.
We identify it as a metastable analog of the Coulomb branch
but now with supersymmetry
broken by the FI terms and vacuum energy $\zeta^2$.  The
metastability corresponds to the possibility for the wrapped
branes to annihilate.
On the other hand, if their space-time separation is large, the dynamics
involving one of the constituents is described by the fractional brane
prescription.  Thus cluster decomposition requires including the
fractional branes.
We develop this argument in detail in section 3 and resolve several
related paradoxes there.

Following \egs, we also claim
that this description is also valid in Matrix theory.
On a qualitative level, the justification is the same --
since the identification of fractional branes with wrapped
branes was justified by BPS arguments, it must survive at strong
string coupling and thus in M theory.
More recently a quantitative prescription for the Matrix theory limit
has been given \senseiberg\ and thus we can check whether these
states survive in this limit.  We do this in section 4,
and find that the states do survive, essentially because they
carry the appropriate D$0$-brane charge for a BPS bound state.

Section 5 contains conclusions.

\newsec{Review of the construction}

We consider type \IIa\ string theory compactified on an ALE $\CM$ asymptotic
to $\BC^2/\Gamma$, or equivalently
M theory on $\CM\times S^1$.  This produces a six-dimensional
theory containing maximal SYM with ADE gauge group $G$, as was first
predicted by considerations of type \IIa\ -- heterotic duality.
The origin of enhanced gauge symmetry is well-known in the language
of eleven-dimensional supergravity -- a two-brane wrapped around the
two-cycle $\Sigma$ of $\CM$ produces a gauge multiplet charged
under $\int_\Sigma C^{(3)}$.  The ALE moduli $\zeta^i$ determine the
complex structure and volume of the two-cycles and translate directly
into the scalars $\phi^i$ of the $d=7$ SYM multiplet.
The expectation values $\int_\Sigma B$ in \IIa\ language become
$\int_\Sigma C^{(3)} = A_{11}$ in M theory language.

In the substringy regime \dkps\ an explicit construction of the ALE
can be given as the moduli space of a D-brane world-volume theory
defined on an orbifold.  This description does not start from
supergravity (rather, gravitational effects emerge) and thus geometric
phenomena such as wrapped two-branes must find new explanations.
However, since supergravity is explicitly described by the closed string
sector, it is not hard to explicitly check these explanations against
the original one.

The Lagrangian is the projection of maximally supersymmetric SYM
with additional FI terms.  We denote coordinates in the ALE as $Z$
and transverse to the ALE (in $\BR^5$) as $X$.
The potential of the gauge theory is
\eqn\pot{
V = \sum [X,Z]^2 + ([Z,Z]-\zeta)^2.
}
As discussed in many places the branch $Z\ne 0$ has ALE topology and
metric.  On this branch the off-diagonal components of $X$ and the gauge
field are massive and thus this world-volume theory has the expected
degrees of freedom of a single D-brane.

If $Z=\zeta=0$ we can give a vacuum expectation value to the off-diagonal
components of $X$.
This is the Coulomb branch, with unbroken $U(1)^n$ gauge symmetry.
The natural interpretation is $n$ D$0$-branes free to move in the
transverse dimensions, each with mass $1/ng_s l_s$.
Each $U(1)$ gauge multiplet contains fermions whose zero modes act on
an $8+8$-dimensional representation space and thus we must identify these
particles as BPS multiplets in space-time.  As described in \egs,
in the orbifold limit, the enhanced $SU(n)$ gauge symmetry
of M theory compactification on the $A_{n-1}$ singularity
is broken by an explicit $B$-field vacuum expectation value, and the only
candidate BPS multiplets are the
massive gauge bosons of this broken symmetry.

The lower dimensional interpretation of the $B$-field vev is as a
Wilson line $A_{11}$ and so this interpretation requires the individual
fractional branes to carry a specific non-zero eleven-dimensional momentum.
In \IIa\ string theory this is RR one-form charge so this can be tested
by world-sheet computation.  Furthermore, the space-time $U(1)^{n-1}$
arises from twisted sector RR fields, so these charges are also computable.

\subsec{World-sheet computation}

We briefly outline the computation of the one-point function of the
relevant RR field vertex operators on a disk whose boundary is a
fractional brane.
For more detail than presented here, we refer to the appendix of \dm\ as
well as to
\ref\has{A. Hashimoto, I.R. Klebanov, Phys.Lett. {\bf B402} (1997) 36,
hep-th/9611214.}.

We first review the world-sheet computation
for flat D-branes.  The RR vertex operator,
in a picture with
superconformal ghost number $-2$ (appropriate for the disk), is
\eqn\RRop{
C_{\mu_1\ldots\mu_n}\left(\Gamma^{\mu_1}\ldots\Gamma^{\mu_n}\right)_
{\alpha\beta}e^{-{3\phi(z)\over 2}}e^{-{\tilde\phi(\bar z)\over 2}}
c(z)\tilde c(\bar z)S^\alpha(z) {\tilde S}^\beta(\bar z).
}
The expectation value on the upper half plane can be computed by
extending the fields on the complex plane using the boundary
conditions
\eqn\doubl{
S^\alpha(z)=\left\{\matrix{
& S^\alpha(z),\hfill &\qquad z\in {\cal H}^+ \cr
& \left(\Gamma^0\ldots\Gamma^p\right)^\alpha_{\dot\beta}{\tilde S}
^{\dot\beta}(z),\hfill &\qquad z\in {\cal H}^-\cr}\right.}
corresponding to $p+1$ Neumann conditions and $9-p$ Dirichlet boundary
conditions. 
This leads straightforwardly to a non-zero expectation value for
the operator corresponding to $C^{(p)}_{0\ldots p}$.

On the orbifold and in the RR sector,
the twist eliminates the fermion zero modes in the internal space,
and the RR boundary state is a bispinor in $d=6$.  For a brane
transverse to the orbifold, this leads to the same expectation value
for any $C^{(p)}$, multiplied by the expectation value of the twist
and Chan-Paton contributions.

The twist part of the vertex operator is simply
$\sigma_g \tilde \sigma_g$ where $\sigma_g$ produces the supersymmetric
ground state in the sector twisted by the group element $g$.
Again by extending the fields,
the one-point function of this operator is $1$.  In correlators involving
other operators the correlation functions with these twist fields can have
cuts and this requires the further association of twist fields with a
contribution to the trace over the boundary Chan-Paton factor \dm.
This leads to the final amplitude
$\CA= \tr \gamma_g$
where $\gamma_g$ is the representation of $g$ on the particular fractional
brane of interest.  For $A_{n-1}$ singularities this will be a phase
$\exp 2\pi i k/n$ for some $k$.

For $g=1$ we find that the charge of a single fractional brane under
the untwisted RR field (and under other untwisted fields, including
the metric and dilaton) is $1/n$ times
the contribution for the original (Higgs branch) D-brane.
The charge for the $g$-twisted RR field is $\gamma_g$ and changing basis
to the twisted RR fields associated with particular nodes of the quiver
diagram leads to a unit charge for the $U(1)$ associated with the
$k$'th node.  We interpret this as a brane wrapped around the two-cycle
whose moduli is controlled by the NS-NS twist fields partner to this RR
field.

Thus an individual fractional brane has the same charge as a wrapped
two-brane and carries $p_{11}=1/n R_{11}$.
Furthermore, the analogous computation for the
Coulomb branch of the original (regular representation) theory
involves a sum of disk diagrams each with Dirichlet boundary conditions
fixed to a particular $X$, so the fractional branes which appear in this
way carry the same charges.

\newsec{Blowing up the two-cycles}

If $\zeta\ne 0$ the target space has ALE topology and metric.
In this situation, the wrapped two-branes are still there, but we
lift the Coulomb branch.  What is their new description?

They must still be the same Coulomb branch, but now no longer supersymmetric
vacua.  This is clear by an adiabatic argument.  One could first separate
the fractional branes with $\zeta=0$, then slowly turn on $\zeta$.
Since they are individually BPS states, they cannot decay.  The resulting
description of branes wrapped around two-cycles of volume $\zeta$ is
the same Coulomb branch, but now lifted to a non-supersymmetric vacuum.
Since we can produce $n$ fractional branes from the vacuum,
space-time cluster decomposition requires us to include the individual
fractional brane sectors.

The effective theory is the same $U(1)^n$ gauge theory, but with a
single additional term obtained by substituting $Z=0$ into the
potential \pot:
\eqn\newL{
\CL = \sum_i \CL_{U(1)_i} + \sum \zeta^2.
}
Although the vacuum is not supersymmetric, the supersymmetry breaking
takes the rather trivial form of a constant shift in the vacuum energy.
This explains how a non-supersymmetric vacuum can still have the requisite
$8n$ fermion zero modes for $n$ BPS multiplets in space-time.

If we turn on individual $X_i$'s to separate the fractional branes, to
a good approximation we can just describe each one by its
individual $\CL_{U(1)_i}$ Lagrangian.  This picture generalizes in an
obvious way to charged states corresponding to non-simple roots and other
bound states.

In general, in these theories with eight supersymmetries
the Coulomb branch metric will get quantum corrections.
These will correspond to long-range forces between the 2B and anti-2B.

All this might be expected, because the state obtained by moving
out on the Coulomb branch has total $G$ gauge charge zero, and one would
be tempted to say that it cannot be BPS.
However, this explanation is not complete, because it suggests
that the Coulomb branch with $\zeta=0$ also should not be a
supersymmetric vacuum.

How can a state containing a wrapped 2B and anti-2B be BPS ?
The answer is that these charges do not add to zero but rather add
to the 0B charge.
Let us write the relevant part of the  charge vector
as $(Q_0,Q_2)$; then (consider the example of an $A_1$ singularity)
the 0B has charge $(1,0)$, and the 2B's have
charge $(1/2,\pm 1)$.
{}From the space-time point of view, the masses of these states are
simply determined in terms of the $d=6$ scalars $\phi$ as
\eqn\masses{
m^2 = \sum_{i=1}^4 (Q_0 \phi^i_0 + Q_2 \phi^i_2)^2.
}
The FI terms correspond to $\phi^i_2$ and as long as these
are zero, all three of the states will be BPS,
with non-zero mass determined by $\phi^i_0 = \int B$
or equivalently the scalar from $A_{11}$.  Call this non-zero
component $\phi^4_0$.
Turning on the FI terms will make the Coulomb branch non-BPS with
$\Delta m^2 \sim \zeta^2$ exactly if they control the three
orthogonal scalars $\phi^i_2$ with $1\le i\le 3$.
This orthogonality can also be seen in the `Narain lattice'
description of the moduli space, motivated by type \II--heterotic duality
\Ab.

Thus the `fractional branes' can also be thought of as D0--D2 bound
states.

\newsec{Matrix theory limit}

In \egs\ it was proposed to define Matrix theory on an ALE using the
same prescription, following the general arguments of \bfss\ which
obtained Matrix theory on flat space by boosting \IIa\ string theory
to the infinite momentum frame.

More recently it has been argued by Sen and by Seiberg that the rescalings
done in a non-systematic way in \bfss\ to derive their Hamiltonian in
fact provide a recipe for producing Matrix theories in general
backgrounds.

The recipe was applied to the case at hand in \hard.
In general all of the structure in a D$0$-brane gauge theory Lagrangian
survives this limit (indeed this scaling was first proposed for this
very reason \kp).  In particular, the form of the ALE metric is
unaffected by the rescaling.

On the other hand, the relation for the other branes is not necessarily
the same.  In particular, the D$2$-branes of toroidal compactification
of \IIa\ theory disappear in the limit, to be replaced by BPS bound states
of the D$0$ and D$2$-branes described by magnetic fluxes in the Matrix
gauge theory.

Are `fractional branes' more like D$0$-branes or like D$2$-branes
from this perspective?
{}From the previous discussion, they are BPS states
carrying both D$0$ and D$2$-brane charges, and like other such
bound states would be expected to survive in Matrix theory.

It was checked in \refs{\egs,\dgm,\dg}\ that the light-cone energy
$P_+$ of these states stays finite in the limit. The ALE Matrix theory is
$U(N)^n$ quantum mechanics with field content specified by an
$A_{n-1}$ quiver diagram, and the careful analysis of the quiver gauge
theory charges in \dg\ leads to the results
\eqn\energy{\eqalign{
P_+&= {nR_{11}\over 2N}{\zeta^2\over (2\pi)^2{l_p}^6}\cr
P_{-} &= {N\over n R_{11}}
}}
We see that $P_+$ is given by the the
expected wrapping contribution to the energy.

\newsec{Conclusions}

We see that the D-brane description of ALE space does contain the
expected BPS states, and realizes the enhanced gauge symmetry of
M theory compactification in the manifest form claimed in \egs.

An interesting problem for future research would be to find
the annihilation cross section for a pair of oppositely
wrapped D$2$-branes to annihilate into a D$0$-brane, a problem
which has not yet been solved for unwrapped D$2$-branes.
A naive extrapolation of the weakly coupled heterotic
string amplitude $O(g_{het})$ would lead to a diverging amplitude
at weak type \II\ coupling, so it seems very likely that this is
a ``BPS algebra structure constant'' \hm\ with explicit dependence on the
moduli.

In this description,
it is determined by the transition amplitude
from the Coulomb to the Higgs branch in a quantum mechanics with
finitely many degrees of freedom.
Given energy $E > \zeta^2$,
the opposite process of pair creation of wrapped branes is classically
allowed \dkps, and it is not at all obvious that such amplitudes will be
suppressed in weak string coupling.

If we compactify Matrix theory on a further $S^1$, we get a $1+1$-dimensional
version of the theory, which has been argued \refs{\wittena}
 to exhibit decoupling
of Coulomb and Higgs branches.

\bigskip
\centerline{Acknowledgments}
We thank P. Aspinwall, S. Lukyanov, J. Maldacena, S. Shenker and
A. Zamolodchikov for discussions.
\bigskip
Note.
This work was largely done in September, as preparation for
studying the question described in the conclusions.
Recently a paper has appeared
which disagrees with the basic premise of this work
\ref\distler{D. Berenstein, R. Corrado, J. Distler, hep-th/9712049.},
prompting us to
publish our results so far.

\listrefs
\end